\documentstyle[12pt]{article}
\newcommand{\be}{\begin{eqnarray}}
\newcommand{\ee}{\end{eqnarray}}
   
\begin{document}
   
\rightline{}
   
\vspace{1cm}
   
\begin{center}
   
   \Large{\bf Tagging nucleon structure functions by heavy particle detection
   in deep inelastic electron-ion scattering}
   \vskip 1cm
   \large{C. Ciofi degli Atti, 
   L.P. Kaptari\footnote{On leave from Bogoliubov Laboratory 
   of Theoretical Physics,
   JINR, 141980, Dubna, Moscow reg., Russia }
   and  S. Scopetta\footnote{At present at Institut f\"ur Kernphysik,
   Universit\"at
   Mainz, Joh.-Joachim-Becher-Weg 45, D-55099 Mainz, Germany.}}
   
   \vspace{1cm}
   
   \footnotesize{\sl
   Department of Physics, University of Perugia and\\
   Istituto Nazionale di Fisica  Nucleare, Sezione di Perugia,\\ Via A.
   Pascoli,
   I-06100 Perugia, Italy\\[3mm]}

   \end{center}
\baselineskip 21pt
   
   \begin{abstract}
   
   It is shown that in deep inelastic electron - ion collisions
   the detection, in coincidence with the scattered electron,
   of a nucleus $A-1$ in the ground state, as well as a nucleon
   and a nucleus $A-2$ also in the ground state,
   may provide unique information on several long standing 
   problems, such as the origin of the EMC effect, the possible
   medium modifications of the nucleon structure functions, and
   the nature of Nucleon-Nucleon correlations.
   \end{abstract}

\vspace{2cm}   

\leftline{PACS: 13.40.-f, 21.60.-n, 24.85.+p, 25.60.Gc}
\leftline{KEYWORDS: semi--inclusive reactions; nucleon structure functions;
medium effects} 

\newpage
   
   At present the possibilities offered by an electron-ion collider
   for investigating the properties of nucleons and nuclei
   are being discussed.
   In view of the capabilities of such a collider to detect
   heavy nuclear fragments resulting from the initial collision
   ~\cite{gsi}, we would like to suggest
   a new kind of semi-inclusive process,
   and illustrate the unique role it could play in clarifying
   several long standing problems, such the origin of the EMC-effect,
   the effect of the nuclear medium on the structure of nucleons, and
   the nature of Nucleon-Nucleon (N-N) correlations. We consider 
   a deep inelastic collision
   of an electron and a nucleus $A$ and propose to measure, in
   coincidence
   with the scattered electron: i) a nucleus $A-1$
   in the ground state; 
   ii) a nucleon and a nucleus
   $A-2$ in the ground state. These semi-inclusive processes radically
   differ
   from the usual 
   ones considered up to now, namely the detection of a nucleon in
   coincidence with the electron~\cite{fs,ciofisim,simula} (only
   in the case of a deuteron target the usual semi-inclusive
   process coincides with process ``i''). In order to make clear
   the physics underlying the above processes we remind few basic concepts
   about
   nucleon momentum distributions $n(k) (k\equiv | \vec k |)$ in the
   parent
   nucleus $A$ and the excitation energy of daughter nuclei $A-1$ and
   $A-2$. The probability to have in the parent a nucleon with momentum
   $k$ and the daughter with
   the excitation energy $E^*_{A-1}$,
   is provided by the
   nuclear spectral function $P^A(k,E)$, $ E=M_{A-1} +M -M_A+ E^*_{A-1}$
   being the nucleon removal energy, i.e. 
   the energy required to remove a nucleon
   from a nucleus $A$ leaving $A-1$ with excitation energy $E^*_{A-1}$
   ($M$ is the nucleon mass).
   
    One has (omitting unecessary here indeces and summations)
   \be
   P^A(k,E) = \label{eq1}
   \sum\limits_{\alpha < F}n^A_\alpha(k)\delta(E-\varepsilon_\alpha)
   +  
       ~ \sum_{f \neq \alpha} \left | \int d \vec{r} ~ e^{i\vec{k} 
   \cdot \vec{r}} ~
       G_{f0}
   (\vec{r}) \right |^2 ~ \delta[E - (E_{A-1}^f - E_A)],
   \ee
   where 
   $E_{A-1}^f=E_{A-1} + E_{A-1}^*$ ($E_{A-1}=M_{A-1}-(A-1)M$),
   $F$ denotes the Fermi energy, $n^A_\alpha(k)$ is the momentum
   distribution
   of a bound shell model state with eigenvalue
   $\varepsilon_\alpha > 0$, and $G_{f0}$ is the overlap between the
   wave functions of the ground state of the parent $A$ and the state $f$
   of the daughter $A-1$, the latter having at least one nucleon in the 
   continuum (see for details ref.~\cite{ciofi-fs,mar,ciofi1}). 
   The first part of the r.h.s. of Eq. (\ref{eq1}) is usually denoted
   $P^A_0(k,E)$, and the second one $P^A_1(k,E)$.
   The so called Momentum Sum Rule links the spectral function to the 
   nucleon momentum distribution, viz.
   \be
   n^A(k)=\int\limits_{E_{\rm min}}^\infty\,P^A(k,E) dE =
   \sum\limits_{\alpha < F}n^A_\alpha(k)
   +  
   ~ \sum_{f \neq \alpha}
   \left | \int d \vec{z} ~ e^{i\vec{k} \cdot
       \vec{z}} ~ G_{f0}
   (\vec{z}) \right |^2~,
   \label{eq2}
   \ee
where $E_{min}=E_{A-1}-E_A$.
   It can therefore be seen that 
$
n^A_0(k) \equiv
\displaystyle\sum\limits_{\alpha
   < F}
   n^A_\alpha(k)= 
\int\limits_{E_{\rm min}}^\infty\,P^A_0(k,E) dE
$ 
represents the momentum distribution in the parent, when
the 
   daughter is either in the ground state or in hole states
   of the parent,
   whereas 
$
n^A_1(k) \equiv n^A(k)-n^A_0(k)=
\int\limits_{E_{\rm min}}^\infty\,P^A_1(k,E) dE
$ 
   represents the momentum
   distribution
   in the parent, when the daughter is left in highly excited states
   with at least one particle in the continuum; this
   means
  that $n^A_0(k)$ is the momentum distribution of weakly (shell model)
   bound nucleons, while $n^A_1(k)$ is the one of deeply bound nucleons
   generated by N-N correlations.
   A realistic model for the latter leads to
   the following form of
   the corresponding spectral function $P^A_1(k,E)$
   \cite{ciofi-fs,mar,ciofi1}
   \be 
   &&
   P^A_1(k,E)= \label{sfcorr}\\[3mm]
   &&
   \int d^3k_{cm} n^A_{rel}\left (|\vec k -\vec k_{cm}/2|\right )
    n^A_{cm} (|\vec k_{cm}|)
   \delta \left [ E -E_{thr}^{(2)} -
   \frac{(A-2)}{2M(A-1)}\cdot \left ( \vec k - \frac{(A-1)\vec
   k_{cm}}{(A-2)}
   \right )^2\right ],
   \nonumber
   \ee
   where $n^A_{rel}$ and $n^A_{cm}$ are, respectively,  
   the relative and Center of Mass
   momentum distributions of a correlated pair. Let us now discuss the two
   processes we are interested in: the first one is the $A(e,e'(A-1))X$
   depicted
   in Fig. 1 (a), where $A-1$ denotes a nucleus
   detected with low momentum and in the
   ground
   state (or in a hole state of the target, the energy interval of these
   states being roughly 20 MeV for a medium weight nucleus). 
   The second process,
   depicted in Fig. 1 (b), is supposed to occur because of N-N
   correlations and therefore, according to eq. (\ref{sfcorr}), implies  
   the detection of a
   nucleon with
   high momentum $\vec p_2$ 
   and a nucleus in the ground or low excited states, with low-momentum 
$
\vec P_{A-2}
\equiv - \vec k_{c.m.}
= -(\vec k + \vec p_2)$ 
($\vec k \equiv \vec p_1$).
   In what follows, kinematics and differential cross sections
   will be given in the ion rest frame; boost trasformations to the
   laboratory system are straightforward.
   The differential cross section 
   for the first process in Impulse Approximation 
   (IA) has the following form  
   
   \begin{eqnarray}
   &&\!\!\!\!\!\! \sigma^A_1\equiv
    \frac{d\sigma^A}{d {\cal E}_k' d\Omega_k \phantom{'\!\!}' d \vec
   P_{A-1}}
   = 
   K^A( x_{Bj},Q^2,y^A) z^A F_2^{N/A}(x_{Bj}/z^A,Q^2)\,n^A_0(|\vec P_{A-1}|),
   \label{crosa-1}
   \end{eqnarray}
   where 
   $\vec P_{A-1} \equiv - \vec p_1$ 
   and
   the kinematical factor $K^A( x_{Bj},Q^2,y^A) $ is
   \begin{eqnarray}
   &&
   K^A( x_{Bj},Q^2,y^A)=
   \frac{4\alpha^2}{Q^6}\,2M {\cal E}_k {\cal E}_k' x_{Bj}\cdot 
     \left( \frac{y}{y^A}\right)^2 
   \left[\frac{(y^{A})^2}{2} + (1-y^A) - 
   \frac{p_1^2x_{Bj}^2(y^A)^2}{z_A^2Q^2}\right ],
   \label{ka}
   \end{eqnarray}
   with  
   $ x_{Bj} = Q^2/2M\nu ; \,\, y=\nu/{\cal E}_k; \,\, Q^2 =-q^2=
  -(k_e-k_e')^2$, 
   $y^A = (p_1\cdot q)/(p_1\cdot k_e)$,
   $z_A = {p_1 \cdot q \over M \nu}
   =p_{10}-|\vec p_1 |\eta \cos\alpha /M$,
   $\eta = |\vec q|/\nu$,
   $ \cos \alpha = \vec p_1 \cdot \vec q / |\vec p_1| |\vec q|$,
   $p_1^0= M_A - 
   \sqrt{ (M_{A-1} + E_{A-1}^*)^2 + \vec P_{A-1}^2}$, and
   $n^A_0(|\vec P_{A-1}|)$ is the momentum distribution of the hit
   nucleon removed from the nucleus $A$ leaving the detected (A-1) nucleus
   in the ground state.
   
   Two issues have to be addressed here: i) the validity of 
   eq. (\ref{crosa-1}), which is based on the IA, and ii) how 
   the process can be used to
   obtain non-trivial information on the 
   nucleon structure functions. To both ends
   we are helped by the fact that $F_2$ and $n^A_0$ are fairly well
   known for the proton and for low values of $| \vec P_{A-1}|$;
   therefore, starting from a target ($Z,N$) and detecting
   a $(Z-1,N)$ ion, the cross section can be related
   to known proton properties. 
   Moreover, it should also be considered that
   $y^A$ and $z_A$ depend very weakly upon $A$, since we are 
   dealing with low momenta and low removal energies 
   ($z_A \sim 1-|\vec p_1| \eta \cos\alpha/M)$. As a result,
   one has $K^A( x_{Bj},Q^2,y^A)\sim K^N( x_{Bj},Q^2,y) =
   ({4\alpha^2}/{Q^6}) 2M {\cal E}_k {\cal E}_k' x_{Bj}
   (y^2/2 + 1 -y -Q^2/4{\cal E}_k^2)$, the relation holding exactly
   in the  Bjorken limit. In order to check the validity of the reaction
   mechanism let us consider the following quantity:
   \begin{eqnarray}
   R(x_{Bj},z_A,z_{A'},\vec P_{A-1},Q^2)& = &\frac{\sigma^A_1}{\sigma^{A'}_1}=
   \frac{K^A}{K_{A'}}
   \frac{z_A F_2^{N/A}(x_{Bj}/z_A,Q^2)}{z_{A'}
   F_2^{N/A'}(x_{Bj}/z_{A'},Q^2)}
   \frac{n_0^A(|\vec P_{A-1}|)}{n_0^{A'}
   (|\vec P_{A-1}|)}
   \cr
   & &
   \longrightarrow\frac{z_A F_2^{N/A}(x_{Bj}/z_A,Q^2)}{z_{A'}
   F_2^{N/A'}(x_{Bj}/z_{A'},Q^2)}
   \frac{n_0^A(|\vec P_{A-1}|)}{n_0^{A'}(|\vec P_{A-1}|)}.\label{eq5}
   \end{eqnarray}
   It can be seen that by investigating the above ratio
   as a function of $|\vec P_{A-1}|$ keeping  $x_{Bj}$ and 
   $\alpha$ fixed (so that $z_A=z_{A'}$),
   one gets
   $R(x_{Bj},z_A,z_{A'},|\vec P_{A-1}|,Q^2)=
    {n_0^A(|\vec P_{A-1}|)}/{n_0^{A'}(|\vec P_{A-1}|)}$, 
   and since for low values of $|\vec P_{A-1}|$ the momentum
   distributions are well known,
   $R(x_{Bj},z_A,z_{A'},|\vec P_{A-1}|,Q^2)$
   can be used
   to check the validity of the spectator mechanism.
   Fig. 2 illustrates the expected behaviour of the 
   ratio for $A=2$, and different
   values of $A'$ 
   (numerically,
   we found that $y^A = y,\, K^A = K^N$,
   $F_2^{N/A}(x_{Bj}/z_A)/F_2^{N/A'}(x_{Bj}/z_{A'}) = 1$
   with an accuracy of few per cents). 
   It can be seen from Fig. 2 that 
   the rapid variation of the ratio
   generated by the variation of $n_0^{A}(|\vec P_{A-1}|)$
   can represent a significant check of the spectator mechanism. 
   The experimental check of the prediction presented in Fig. 2
   is a prerequisite for any further
   measurements, since any deviation from this prediction
   represents strong indications of some drawbacks of the spectator model.
   
   Let us now consider the possibility to investigate the nucleon 
   structure functions; 
   this has been recently addressed in ref.~\cite{simula},
   where the ratio  $D(e,e'p)X/D(e,e'n)X$ has been considered in 
   order to investigate the neutron 
   to proton structure function ratio. Here we are interested
   in emphasizing the origin of the EMC-effect and
   possible medium modifications of the nucleon
   structure functions. To this end in Fig. 3 we show
   the ratio
   ~(\ref{eq5}) vs. $x_{Bj}$ for 
   $A'=2$ and various values of $A$, calculated at
   fixed value of $|\vec P_{A-1}|$
   chosen such that $z_D = z_A$. 
   In such a way the 
    nucleon structure functions 
   depend only upon $x_{Bj}$ and the ratio is a constant (curves
   (a))
   with the absolute value given by the ratio of the momentum
   distributions.
   On the contrary,
   if we consider the $Q^2$-rescaling model~\cite{close}
   with $Q^2_A= \xi_A(Q^2) Q^2$, 
   we get an $A$-dependent ratio having an EMC-like behaviour
   (curves (b)). Therefore, one may conclude that these processes
   may serve as a tool to distinguish various interpretations of the 
   EMC-effect. In order to better emphasize the difference 
   between $x$-rescaling and $Q^2$-rescaling models, we have calculated
   the same ratio at large values of $|\vec P_{A-1}|$ and not such that 
   $z_A=z_D$; the latter condition leaves unaffected $Q^2$-rescaling
   but it does affect the $x$-rescaling since the ratio will now depend
   upon $x_A = x_{Bj}/z_A$, with $x_D > x_A$, because of the important role
   of the kinetic energy in the definition of $z_D$ 
   ( $ z_D \approx (1 - p_1^2/2M^2) < z_A$). This effect is clearly seen
   in Fig. 4,
   where the ratio 
   calculated within the $x$-rescaling model is predicted to
   increase, whereas the $Q^2$-rescaling model gives again 
   the EMC-like behaviour.
     
   In closing we would like to stress that
   the semi-inclusive processes on weakly bound nucleons 
   that we have analyzed, can be used
   to investigate the ratio of the neutron to proton structure functions
   by performing experiments $A(e,e'A-1)X$ on mirror nuclei or
   $A(e,e'N-1)X$ and $A(e,e'Z-1)X$ on the same isoscalar target.

   We address now the important issue of a possible medium modification 
   of the nucleon structure functions. The semi-inclusive process offers
   the possibility to investigate the nucleon structure function 
   for weakly and deeply bound nucleons separately. To this end one has to
   consider the process $A(e,e' N_2 (A-2))X$ depicted in Fig. 1 (b).

   The differential cross section of such a process in IA
   reads as follows

   \begin{eqnarray}
   \sigma_2 & \equiv &  \frac{d\sigma^A}{d {\cal E}_k'
   d\Omega_k\phantom{'\!\!}' 
   d \vec P_{A-2} d \vec p_{2}}  =
   \cr
   & & 
    K^A( x_{Bj},Q^2,y^A) 
    \cdot  z_A  F_2^{N/A}(x_{Bj}/z_A,Q^2)
   \,n^A_{cm}(|\vec P_{A-2}|)\, n^A_{rel.}(|\vec p_2 +\vec P_{A-2}/2 |) ,  
   \label{crosa-2}
   \ee
   where the notations are the same as in 
   the previous process. In spite of the fact that
   eqs.~(\ref{crosa-1}) and (\ref{crosa-2}) 
   are very similar, the underlying physics is completely different,
   since
   the nucleon structure function $F_2^{N/A}(x_{Bj}/z_A,Q^2)$ 
   in the first case 
   represents 
   the quark distribution in a nucleon which is
   almost free, while in the second
   case the hit nucleon is strongly bound in the parent nucleus and
   its structure function may be affected by the
   so-called off-mass-shell deformations
   (see, for instance refs. \cite{mthomas,kukhanna}). 
   As a matter of facts, if 
   the nucleon structure function could be
   extracted from the cross section 
   (\ref{crosa-2}) and compared 
   with the one obtained 
   from  the cross section (4),
   a direct comparison of nucleon structure functions
   for weakly and deeply bound nucleons can, for the first time,
   be carried out.
   
   It should be pointed out that,
   since kinematically $y^A$ is connected to high momenta $|\vec p_2|$,
   the factor $ K^A(x_{Bj},Q^2,y^A)$ may strongly
   differ from $ K^N(x_{Bj},Q^2,y)$,
   unless proper kinematical conditions 
   are chosen, which turn out to be small values of $x_{Bj}$ or
   the Bjorken limit, $Q^2\to\infty$.
   Nevertheless,
   we will now show that, in this process, 
   it is still possible to investigate separately
   the momentum distribution and the structure functions of strongly
   bound nucleons. First of all we choose a combination of kinematical 
   variables such as to assure that $K^A=K^N$.
   We found that  
   for  $Q^2=20$ GeV$^2$ and $x_{Bj}=0.05$,
   the direction of the transfered momentum 
   $\vec q$
   coincides, in the frame where the target
   is at rest, with the electron beam direction
   ($ \theta_{\widehat{kq}}\approx 2^0$); in this case,
   $y^A\simeq y$ and
   $K^A \simeq K^N$
   (our numerical estimates
   show that 
   $K^A/K^N$
   varies from $0.99$ at $|\vec p_2|=350$ MeV to $0.96$ at 
   $|\vec p_2|=1$ GeV);
   in the laboratory system,
   adopting realistic figures for a possible collider, i.e.
   ${\cal E}_k \approx 5 $ GeV, $T_N$ = (kinetic
   energy per nucleon) $\approx 25$ GeV \cite{gsi},
   the chosen values of $Q^2$ and $x_{Bj}$ correspond
   to 
   ${\cal E}_k'\approx 2\,GeV;\quad \theta_{\widehat{kk'}} \approx 90^0$
   (in the ion rest frame they
   correspond to ${\cal E}_k\sim 260$ GeV, 
   $ {\cal E}_k'\approx 50$ GeV; $\theta_{\widehat{kk'}} \approx 2^0$).

   The validity of eq. (\ref{crosa-2}) can now be tested
   in the following way.
   We will take advantage of the observation ~\cite{pan} that for high 
   values of $|\vec p_1|$ 
   the nucleon
   momentum distribution for a complex nucleus turns out
   to be the rescaled momentum distribution of the deuteron
   with very small $A$ dependence (unlike what happens
   for the low momentum part of $n(k)$ (cf. Fig. 2)). Let us
   consider the following ratio:
   \begin{eqnarray}
   R(x_{Bj},z_A,z_{A'},Q^2,\vec P_{A-2},\vec p_2)\, & \equiv\,& 
   \displaystyle\frac{\sigma_2^A(x_{Bj},Q^2,\vec P_{A-2},|\vec p_2|)}
   {\sigma_2^{A'}(x_{Bj},Q^2,\vec P_{A-2},|\vec p_2|)}
   \cr
   & = & 
   \frac{z_A}{z_{A'}}
   \frac{F_2^{N/A}(x_{Bj}/z_A,Q^2)}{F_2^{N/A'}(x_{Bj}/z_{A'},Q^2)}
   \cdot
   \frac{n^A_{rel.}(p_{rel.})}{n^{A'}_{rel.}(p_{rel.})}~,
   \label{ratioa}
   \end{eqnarray}
   where $p_{rel}=|\vec p_2 +\vec P_{A-2}/2 |$.
   Let us keep fixed $|\vec P_{A-2}|$; then, if
   we could make $F_2^{N/A}\simeq F_2^{N/A'}$, the ratio,
   measured at $p_{rel} \geq  2 - 3$ $f^{-1}$, would be a constant,
   since $n^A_{rel.}\propto
   n^{D}$ for any value of $A$.
   The condition $F_2^{N/A}\simeq F_2^{N/A'}$ can be fulfilled by properly
   choosing $\vec p_2$ and $\vec P_{A-2}$ for, if Eq. (3)
   is correct, the off-shell energy of the hit nucleon
   is 
   $p_1^0= M_A - 
   \sqrt{(M_{A-1} + E_{A-1}^*)^2 + \vec p_1^{\,\,2} }$, 
   with $E_{A-1}^* \simeq
   \frac{(A-2)}{2M(A-1)} |\vec p_1|^2$,
   so that 
   $z_A \simeq 1 - {E^A_{min} \over 2 M} - {(A-2) |\vec p_1|^2 
   \over 2 M^2 (A-1)} - 
   {|\vec p_1| \over M} \cos \alpha$, and by properly choosing
   $\vec p_2$ and $\vec P_{A-2}$ ($\vec p_1 = -(\vec p_2 
   + \vec P_{A-2})$) one can make $z_A \simeq z_{A'}$, i.e.
   $F_2^{N/A} \simeq F_2^{N/A'}$ (note that for large values
   of $| \vec p_1|$, as it is in our case, and large values of $A$,
   the dependence of $z_A$ upon $A$ is unessential).
   Another possibility would be to consider in Eq. (6) $A=A'=2$,
   with the denominator taken at some fixed and high value
   of $|\vec P_{A-1}|$ = $|\vec {p_1}|$ = $ \tilde p $, so that
   ${\tilde z_{D'}} \simeq 1 - {E_D \over M} - {\tilde p^2 \over 2 M^2} - 
   {\tilde p \over M} \cos \alpha$. If we now consider
   Eq. (8) with $A=A'>2$ and the denominator taken at the value
   $|\vec p_1| = |\vec p_2 + \vec P_{A-1}| = \tilde p$, so that 
   $\tilde z_{A'} \simeq 1 - {E^A_{min} \over M} - 
   {(A-2)\tilde p^2 \over 2 M^2(A-1)} - 
   {\tilde p \over M} \cos \alpha$, the ratio (8) plotted
   versus $p_{rel}$ for a fixed value of 
   $\vec P_{A-2}$ should behave as the ratio (6).
     
    If such a deuteron -- like
    behaviour of Eq. (\ref{ratioa})
    is found, this would represent a stringent test
    of the spectator model, and off-mass-shell effects
    can be investigated by
    comparing the structure functions measured in the $A-1$ and
    $A-2$ processes.
    To this end:
   
    i) in the case of (A-1) one may fix $|\vec P_{A-1}| = 50$ MeV
       (weakly bound nucleon), 
    $Q^2 = 20$ GeV$^2$
    and
    vary $x_{Bj}$ from $0$ to $0.5$ and the angle of  $\vec P_{A-1}$
    from $0^o$ to $180^o$. One finds that $K^A=K^N$ within 
   $\pm 3\%$. This condition allows one to
   vary  the argument of the nucleon structure function 
    from  $0$ to $0.6$ and to study $F_2^{N}$ in this interval. Note
   that in  order to exclude the influence of the momentum distribution,
   one may  vary any variables but $|\vec P_{A-1}|$;
   
   ii) in the case of (A-2), one can choose 
   $|\vec P_{A-2}| =50$ MeV, $\,|\vec p_2| = 700$
   MeV (deeply bound nucleon),
   the angle  between them constant, e.g. $180^o$, 
   $\, Q^2 = 20$ GeV$^2$ and
   $x_{Bj}=0.05$. Then varying the angle between $\vec p_2$ and 
   $\vec q$ from zero to the kinematical limit
   $x_{Bj}/z = 1$, one may obtain the structure
   function of a strongly bound nucleon in the whole interval of $x$.
   A detailed treatement of the $A(e,e' N_2 (A-2))X$ process with
partial account of Final State Interactions due to both the 
$N_2 - (A-2)$ rescattering and fragmentation of the hit nucleon
will be presented elsewhere.
\vskip 6mm
\leftline{\Large{\bf Acknowledgements}}
\vskip 4mm
\noindent
One of us (C.C.d.A.) would like to thank 
V. Metag, D. v Harrach and A. Sch\"afer for useful discussions during
the Working Groups on ``Long Term Perspectives of GSI".

\newpage

\leftline{\Large{\bf Captions}}

\vskip 1cm
\noindent
{\bf Figure 1}: Kinematics 
for the semi-inclusive processes in the impulse approximation.
\vskip 5mm
\noindent
{\bf Figure 2}:
The ratio
$ R\,(x_{Bj},z_D,z_{A},| \vec P_{A-1}|,Q^2) 
   \equiv \displaystyle\frac{\sigma_1(D)}{\sigma_1 (A)} 
   = \displaystyle
    \frac{z_D}{z_A}\frac{n_D(|\vec P_{A-1}|)}{n_A(|\vec P_{A-1}|)}
    \cdot
\displaystyle\frac{F_2^{N/D}(x_{Bj}/z_D,Q^2)}{F_2^{N/A}(x_{Bj}/z_A,Q^2)}$,
   (Eq. (6))
   calculated for different values of $A$ at fixed
   $x_{Bj}$ and
   $\alpha = 90^0$; 
   because of the latter condition, 
   $z_D$ differs from $z_A$ at low values of
   $|\vec P_{A-1}|$ by only few percents, so that 
   the behaviour of the curves
   is given by the ratio 
   ${n_D(|\vec P_{A-1}|)}/{n_A(|\vec P_{A-1}|)}$.
   The nucleon momentum distributions 
   have been taken from ref.~\protect\cite{ciofi1}.
\vskip 5mm
\noindent
{\bf Figure 3}:
The ratio
   $ R(x_{Bj},z_A,z_{A'},| \vec P_{A-1}|,Q^2) 
   \equiv \displaystyle\frac{\sigma_1(A)}{\sigma_1(D)} 
   = \frac{z_A}{z_D}\displaystyle\frac{n_A(|\vec P_{A-1}|)}{n_D(|\vec P_{A-1}|)}
    \cdot 
   \displaystyle\frac{F_2^{N/A}(x_{Bj}/z_A,Q^2)}
   {F_2^{N/D}(x_{Bj}/z_D,Q^2)}$, for different values of $A$
   at $\alpha = 90^0$.  
   For each nucleus $|\vec P_{A-1}|$ is chosen such that $z_D=z_A$. Curves labeled
   (a) correspond to x - rescaling
   while those labeled (b) 
  correspond to
   $Q^2$-rescaling~\protect\cite{close}.
   The behaviour of the curves is given by the ratio 
   $F_2^{N/A}(x_{Bj},Q^2)/F_2^{N/D}(x_{Bj},Q^2)$.
\vskip 5mm
\noindent
{\bf Figure 4}:
The same ratio as in Fig. 3 but
   $|\vec P_{A-1}|$ is the same for each target,
   so that the behaviour is governed by the ratio
   $F_2^{N/A}(x_{Bj}/z_A,Q^2)/F_2^{N/D}(x_{Bj}/z_D,Q^2)$.
   Since the hit nucleon is an almost free one,
   $z_A \sim 1-{E_{min}^A \over M}\sim 1$, whereas $z_D \simeq
   1-{E_D \over M} - {p^2 \over 2 M^2} <1$, therefore in 
   x - rescaling the ratio is predicted to increase as $x_{Bj}$ increases
   (curves (a)),
   whereas in $Q^2$ rescaling model~\protect\cite{close} this ratio
   has a decreasing behaviour (curves labeled (b)).

\newpage

   \begin{figure}[ht]
    \let\picnaturalsize=N
   \def\picsize{8cm}
   \def\picfilename{one_a.eps}
   \ifx\nopictures Y\else{\ifx\epsfloaded Y\else\input epsf \fi
   \let\epsfloaded=Y
   \centerline{\ifx\picnaturalsize N\epsfxsize
    \picsize \epsfysize 8cm
    \fi \epsfbox{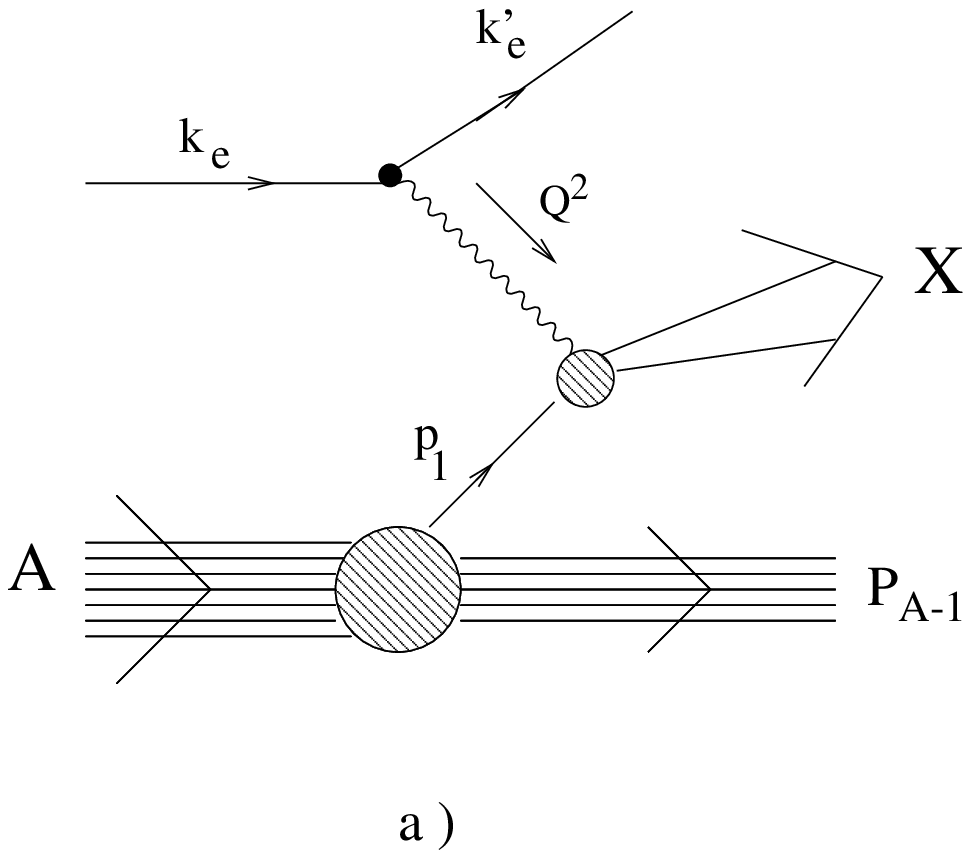} \epsfysize 7.5cm \epsfbox{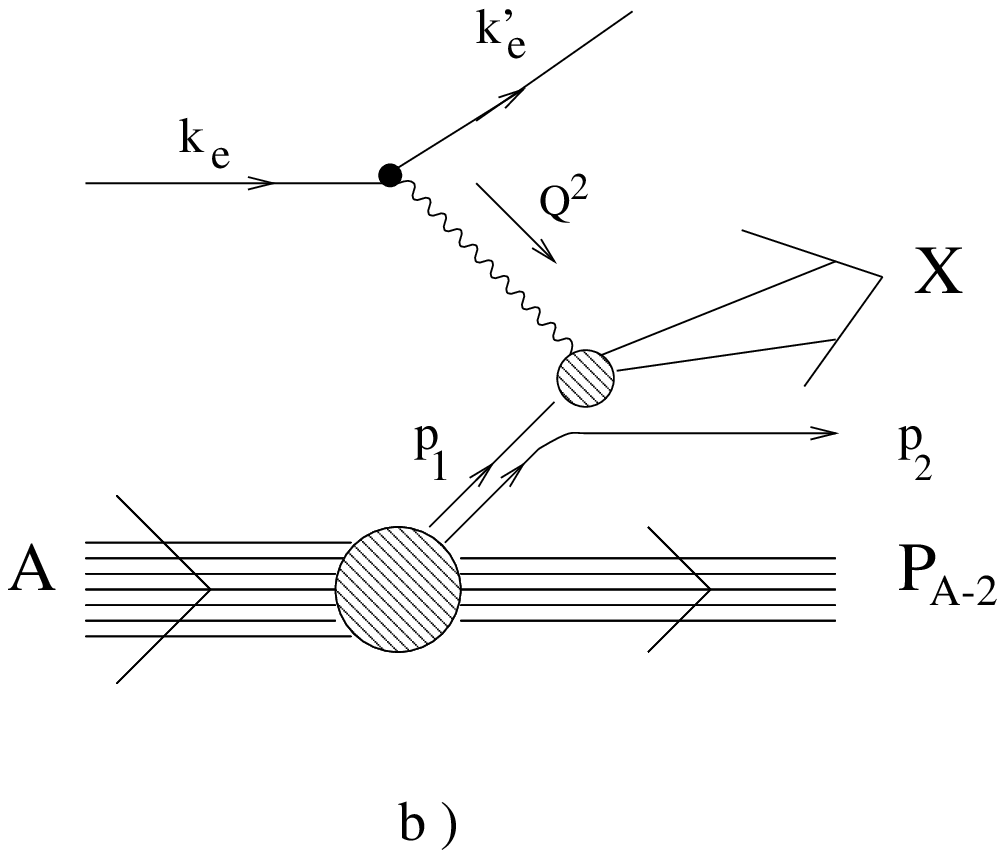}}}\fi
   \protect\label{diagram}
  \end{figure}
\begin{center}
{\Large C. Ciofi degli Atti, Phys. Lett. B}
\\[4mm]
{\Large{\bf FIGURE 1}}
\end{center}   
   
   \newpage
   \vspace*{2cm}
   \begin{figure}[[ht]
    \let\picnaturalsize=N
   \def\picsize{16cm}
   \def\picfilename{two.ps}
   \ifx\nopictures Y\else{\ifx\epsfloaded Y\else\input epsf \fi
   \let\epsfloaded=Y
   \centerline{\ifx\picnaturalsize N\epsfxsize
    \picsize \epsfysize 24cm\epsfxsize 12cm
    \fi \epsfbox{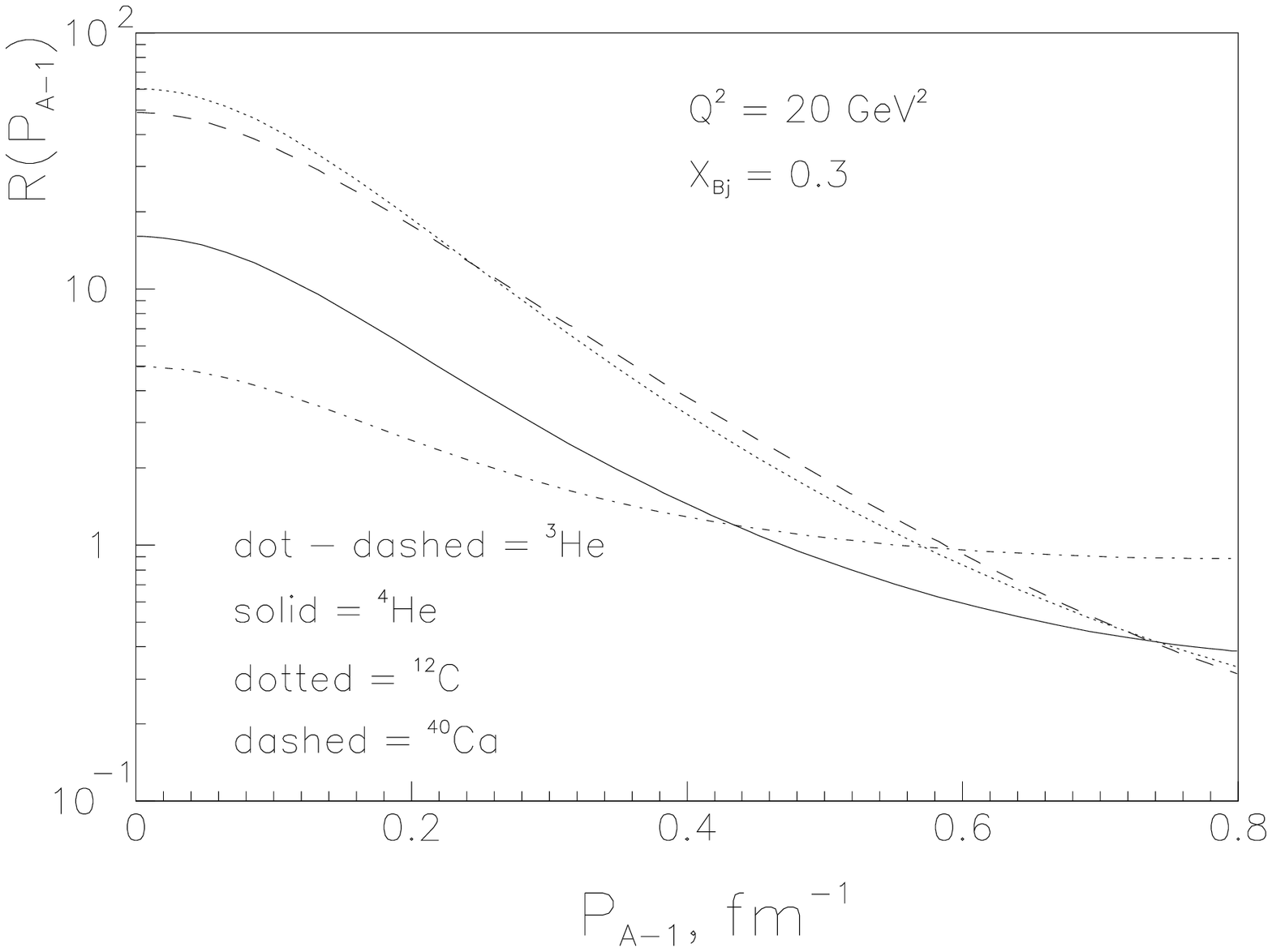}}}\fi
   \vspace*{-4cm}
   \protect\label{mdistr}
    \end{figure}
\vspace*{-3cm}
\begin{center}
{\Large C. Ciofi degli Atti, Phys. Lett. B}
\\[4mm]
{\Large{\bf FIGURE 2}}
\end{center}

   \newpage
   \vspace*{2cm}

   \begin{figure}[[ht]
    \let\picnaturalsize=N
   \def\picsize{12cm}
   \def\picfilename{three.ps}
   \ifx\nopictures Y\else{\ifx\epsfloaded Y\else\input epsf \fi
   \let\epsfloaded=Y
   \centerline{\ifx\picnaturalsize N\epsfxsize
    \picsize \epsfysize 24cm\epsfxsize 12cm
    \fi \epsfbox{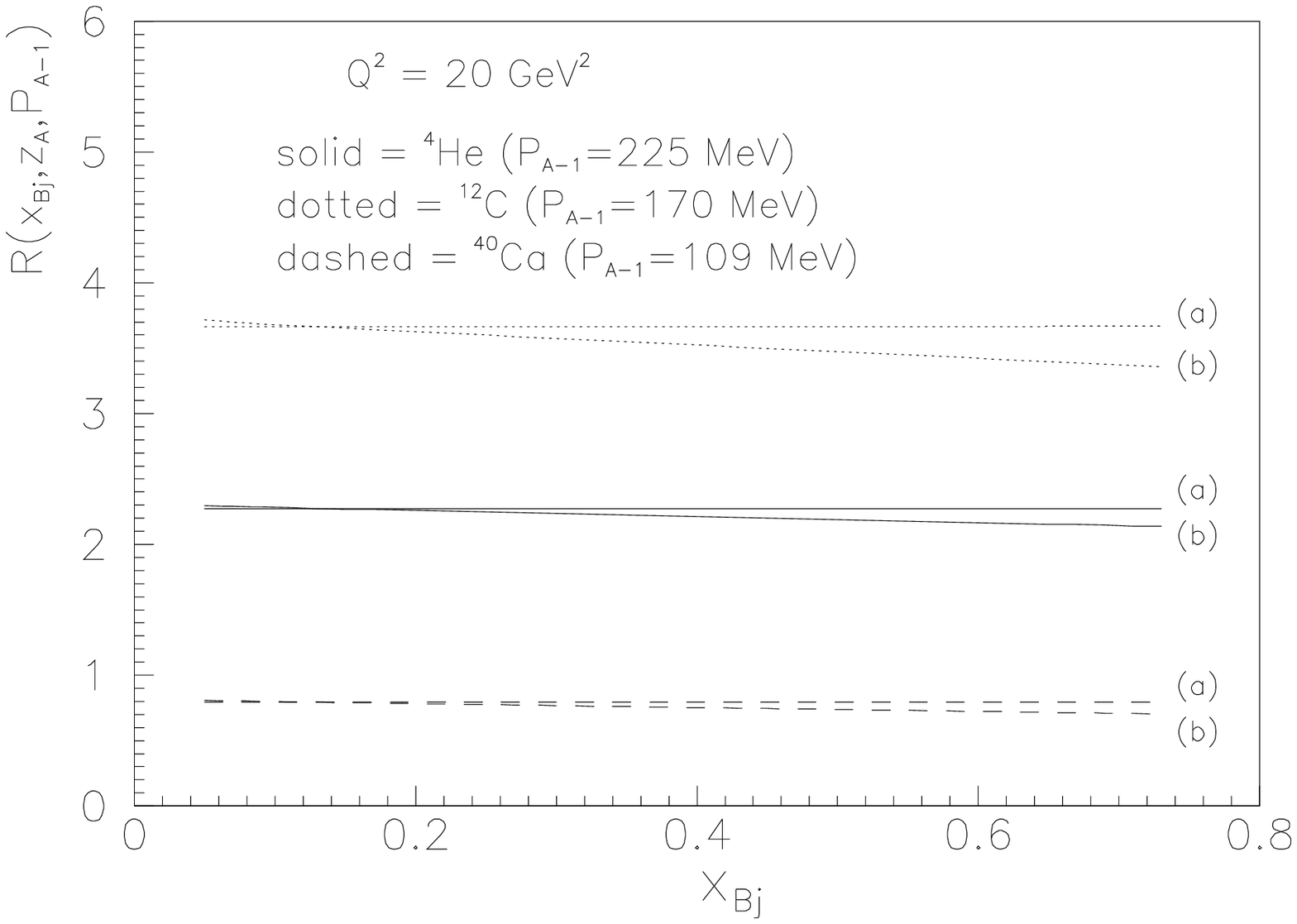}}}\fi
   \vspace*{-4cm}
   \label{rescaling}
    \end{figure}
\vspace*{-3cm}
\begin{center}
{\Large C. Ciofi degli Atti, Phys. Lett. B}
\\[4mm]
{\Large{\bf FIGURE 3}}
\end{center}

   \newpage
   \vspace*{2cm}

   \begin{figure}[[ht]
     \let\picnaturalsize=N
   \def\picsize{16cm}
   \def\picfilename{four.ps}
   \ifx\nopictures Y\else{\ifx\epsfloaded Y\else\input epsf \fi
   \let\epsfloaded=Y
   \centerline{\ifx\picnaturalsize N\epsfxsize
    \picsize \epsfysize 24cm \epsfxsize 12cm
    \fi \epsfbox{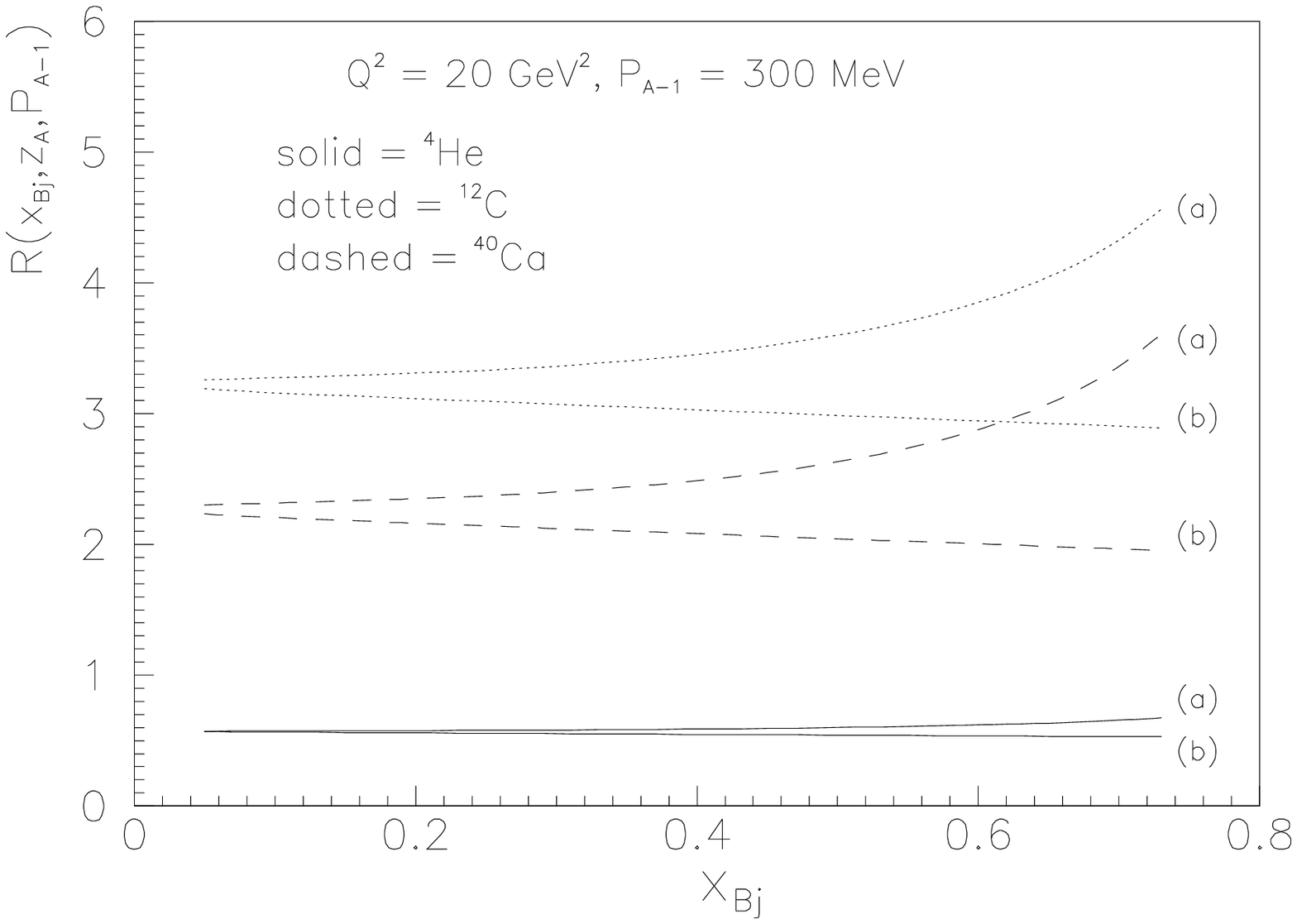}}}\fi
   \vspace*{-4cm}
   \protect\label{res300}
    \end{figure}   
\vspace*{-3cm}
\begin{center}
{\Large C. Ciofi degli Atti, Phys. Lett. B}
\\[4mm]
{\Large{\bf FIGURE 4}}
\end{center}

\end{document}